\documentstyle[12pt,a4,epsf]{article}
\begin{document} 
\bibliographystyle{plain} 
\input{epsf} 
{\Large \bf \noindent Langevin Equation for the Density of a System of
 Interacting Langevin Processes} 
\vskip 2 truecm
\noindent David S. Dean   
\pagestyle{empty} 
\vskip 0.5 truecm 
\vskip 1 truecm
\noindent Dipartmento di Fisica and INFN, Universit\`a di Roma, La Sapienza, 
P. A. Moro 2, 00185 Roma (Italy)
\vskip 1 truecm 
\noindent{\bf PACS:} 
\vskip 1 truecm \noindent{\bf Abstract:} 
We present a simple derivation of the stochastic equation obeyed by the density
function for a system of Langevin processes interacting via a pairwise 
potential. The resulting equation is considerably different from the
phenomenological equations usually used to describe the dynamics of non conserved
(Model A) and conserved (Model B) particle systems. The major feature is that
the spatial white noise for this system appears not additively but 
multiplicatively. This simply expresses the fact that the density cannot 
fluctuate in regions devoid of particles. The steady state for the density
function may however still be recovered formally as a functional integral
over the coursed grained free energy of the system as in Models A and B. 
\vskip 2 truecm 
\noindent  July 1996
\newpage
\pagenumbering{arabic} 
\pagestyle{plain} 
\def\half{{1\over 2}}
\def\OO{\Omega}
\def\sech{{\rm sech}}
\def\n{{\newline}}
 \def\aa{\alpha}
 \def\bk{{\bf k}}
 \def\bkp{{\bf k'}}
 \def\bqp{{\bf q'}}
 \def\bq {{\bf q}}
 \def\EE{\Bbb E}
 \def\EEx{\Bbb E^x}
 \def\EEo{\Bbb E^0}
 \def\LL{\Lambda}
 \def\PP{\Bbb P^o}
 \def\rr{\rho}
 \def\SS{\Sigma}
 \def\ss{\sigma}
 \def\ll{\lambda}
 \def\dd{\delta}
 \def\ww{\omega}
 \def\ll{\lambda}
 \def\DD{\Delta}
 \def\DDt{\tilde {\Delta}}
 \def\kr{\kappa\lb \LL\rb}
 \def\PPx{\Bbb P^{x}}
 \def\gg{\gamma}
 \def\kk{\kappa}
 \def\tt{\theta}
 \def\bs{\hbox{{\bf s}}}
 \def\bh{\hbox{{\bf h}}}
 \def\lb{\left(}
 \def\rb{\right)}
 \def\prt{\tilde p}
\def\pt{\tilde {\phi}}
 \def\bb{\beta}
 \def\hal{{1\over 2}\nabla ^2}
 \def\bg{{\bf g}}
 \def\bx{{\bf x}}
 \def\bu{{\bf u}}
 \def\by{{\bf y}}
 \def\hag{{1\over 2}\nabla}
 \def\beq{\begin{equation}}
 \def\eeq{\end{equation}}
 \def\cosech{\hbox{cosech}}

Recently there has been a tentative attempt \cite{MePa} to analyse the liquid--glass transition within the context of the replica approach to disordered 
systems. A program of implementing replica techniques from disordered systems
has had some success in capturing the properties of systems without disorder
but which have many metastable states \cite{BoMe, MaPaRi1, MaPaRi2}.
 As the glass 
transition is essentially dynamical it is useful to have a simple dynamical
model for these types of problems. In the literature on liquids there has 
been a considerable effort to analyse the dynamics of liquid systems. In this
letter we present a simple microscopic derivation for the evolution of the
of the density of a system of Langevin processes interacting via pairwise 
potentials. Rather by working with Fokker--Planck equations in large coordinate
 spaces and then using projection operator techniques to obtain a hierarchy
of equations for the $n$ point density correlation functions we simply derive
a closed functional Langevin equation for the evolution of the density. Using
this it is the trivial to derive the dynamical BBGKY 
(e.g. see \cite{HaMc, SoKrSa}  and references therein) hierarchy of equations.
Interestingly the naive course grained free energy of the system appears quite
naturally. However the equation is not amongst the class of equations used
often to describe the approach to equilibrium of systems specified by their
free energy as a function of some scalar order parameter \cite{Hal,Bra}.

We commence by considering the dynamics of a system of particles in a thermal
white noise heat bath. The particles interact via a pairwise potential $V(x)$, 
this could be a Coulomb type of interaction, a Leonard--Jones type interaction
or simply a hard sphere repulsion. Each particle experiences a thermal white
noise and moves under the force generated from the potential due to its
neighbours. Consequently the $i$th particle obeys the Langevin equation 
\beq {dX_i(t)\over dt} = \eta_i(t) - \sum_{j = 1}^N \nabla V(X_i(t) - X_j(t)), \eeq
to ease notation we have assumed a potential such that $\nabla V(0) = 0$.
The noise is uncorrelated in time and the noise acting on a single particle is
not correlated with the noise acting on the others. The components of the noise
are also taken to be uncorrelated, hence:
\beq \langle \eta_i^\mu(t) \eta_j^\nu(t')\rangle = 2T\delta_{ij}\delta^{\mu\nu}\delta(t-t').\eeq
Our strategy will be to consider the evolution of the density function for a 
single particle
\beq \rho_i(x,t) = \delta(X_i(t) - x), \eeq
we shall then demonstrate how one may write a closed Langevin equation for the
global density
\beq \rho(x,t) = \sum_{i=1}^N \rho_i(x,t) .\eeq
The derivation follows a well known argument. Consider an arbitrary function $f$
defined on the coordinate space of the system. Using the definition of the
density it is a tautology that
\beq f(X_i(t)) = \int dx \rho_i(x,t)f(x). \label{eq:prf0}\eeq
Expanding the stochastic differential equation using the Ito calculus
(for example see \cite{Oks}) over the next time step $\delta t$ one obtains
\beq {df(X_i)\over dt}= \int dx \rho_i(x,t)\lb \nabla f(x)\cdot \eta_i(t) -
\nabla f(x) \cdot \lb \sum_{j= 1}^N \nabla V(x - X_j(t)) \rb + T\nabla^2 f(x) \rb. 
\eeq
Rearranging the above integral by integration by parts we obtain
\beq {df(X_i)\over dt}= \int dx f(x)\lb - \nabla \cdot(\rho_i(x,t) \eta_i(t)) +
\nabla \cdot\rho_i(x,y) \lb \sum_{j= 1}^N \nabla V(x - X_j(t)) \rb + T\nabla^2 
\rho_i(x,t) \rb \label{eq:prf1}.\eeq
However from (\ref{eq:prf0}) we may also deduce
\beq {df(X_i)\over dt}= \int dx {\partial\rho_i(x,t)\over \partial t} f(x)
\label{eq:prf2}.\eeq
Comparing equations (\ref{eq:prf1}) and (\ref{eq:prf2}) we find 
(using the fact that $f$ is an arbitrary function) that
\beq {\partial\rho_i(x,t)\over \partial t} = - \nabla \cdot(\rho_i(x,t) \eta_i(t)) +
\nabla \cdot\lb\rho_i(x,t)  \sum_{j= 1}^N \nabla V(x - X_j(t)) \rb + T\nabla^2 
\rho_i(x,t). \label{eq:prf3} \eeq
We emphasise that this argument is standard and the only subtlety is
 that we have not carried out any thermal averaging at this point.
Summing equation (\ref{eq:prf3}) over the $i$ and using the definition of the 
density $\rho$ we obtain
\beq {\partial\rho(x,t)\over \partial t} = -\sum_{i = 1}^N \nabla 
\cdot(\rho_i(x,t) \eta_i(t)) +
\nabla \cdot\lb\rho(x,t)\int dy\rho(y,t) \nabla V(x - y)\rb  + T\nabla^2 
\rho(x,t) .\label{eq:prf5}\eeq
This equation is {\em almost} a closed equation for $\rho$, the problem is that
the noise term appears to contain too much information  about the individual
$\rho_i$ and hence, in the current form, it is not a Markovian equation for the
evolution of the global density. We define this noise term to be  
\beq \xi(x,t) = -\sum_{i=1}^N \nabla\cdot(\eta_i(t) \rho_i(x,t) ), \eeq
it is clearly still Gaussian and the correlation function is given by
\beq \langle \xi(x,t)\xi(y,t') \rangle = 2T
 \delta(t-t')\sum_{i=1}^N\nabla_x \cdot 
\nabla_y\lb \rho_i(x,t)\rho_i(y,t)\rb. \eeq
However the above term simplifies using a trivial property of the Dirac delta 
function, explicitly
\beq \rho_i(x,t)\rho_i(y,t) = \delta(x-y)\rho_i(x,t) = \delta(x-y)\rho_i(y,t)
.\eeq
Using this one finds
\beq \langle \xi(x,t)\xi(y,t') \rangle = 2T\delta(t-t')
\nabla_x \cdot\nabla_y \lb\delta(x-y)\rho(x,t)\rb. \eeq
At this point we shall consider redefining the white noise by introducing a 
global noise field
\beq \xi'(x,t) = \nabla\cdot\lb\eta(x,t)\rho^{\half}(x,t)\rb, \eeq
where $\eta$ is a global uncorrelated white  noise field such that
\beq \langle \eta^\mu(x,t)\eta^\nu(y,t') \rangle =  2T\delta(t-t')\delta^{\mu\nu}\delta(x-y). \eeq
It is easy to see that both Gaussian noises $\xi$ and $\xi'$ have the same
correlation functions and are therefore statistically identical. We may 
therefore write equation (\ref{eq:prf5}) as
\beq {\partial\rho(x,t)\over \partial t} =
 \nabla\cdot\lb \eta(x,t)\rho^{\half}(x,t)\rb  +
\nabla \cdot\lb\rho(x,t) \int dy \rho(y,t) \nabla V(x - y)\rb  + T\nabla^2 
\rho(x,t) \label{eq:prf6}. \eeq
Equation (\ref{eq:prf6}) is clearly a closed
 Langevin equation for the evolution
of the particle density of the system. The crucial point is that on insisting 
on knowing the global property of the white noise field we have gained 
sufficient information to write down a Markovian evolution equation for $\rho$.
If one were to consider the course grained free energy for the system one would
be lead to write
\beq F = \half \int dx dy \rho(x)V(x-y)\rho(y) + T\int dx \rho(x) \log\rho(x) 
,\eeq
up to constant terms in the case of a fixed particle number.
Using this as a definition of $F$ it is easy to see that (\ref{eq:prf6}) may
 be written as 
\beq {\partial \rho(x,t)\over \partial t} =  \nabla \cdot \lb \rho(x,t) 
\nabla {\delta F\over \delta \rho(x)}\vert_{\rho(x,t)} \rb + \nabla\cdot\lb \eta(x,t)\rho^{\half}(x,t)\rb. \label{eq:prf7} \eeq
Two things are immediately obvious from equation (\ref{eq:prf7}), firstly
particle number is conserved and secondly in regions where the system is void
of particles the density cannot change and hence stays zero. The above equation
is in fact a special case of an equation derived semi--phenomenologically by
Kawasaki some 20 years ago in the context of non--polymeric incompressible 
binary fluids \cite{KaKo}.
As is usual in such derivations, the free energy is taken to
be of the Ginsburg--Landau type directly, if one sets the velocity field to 
zero one obtains formally the same equation as (\ref{eq:prf6}). The derivation in the case of zero velocity field is brief enough to be included here for the
sake of completeness. One assume that the density obeys an equation of the form
 \beq {\partial \rho \over \partial t} = -\nabla \cdot j +
 \xi(x,t), \eeq
where $j$ is the current and $\xi$ is  the divergence of a 
 random force uncorrelated in time.
In the course grained microscopic approach $j$ is simply given by
\beq j = -\rho\nabla{\delta F \over \delta \rho}, \eeq
the fluctuation dissipation theorem (see for example the discussion in
 \cite{Lan}) then requires that the noise $\xi$ must
have exactly the same correlation function as the noise in (\ref{eq:prf6}),
hence the result.

Here we
briefly recall the two standard models used in dynamical analysis of phase
ordering kinetics and other areas \cite{Hal,Bra}.
 The basic approach here has to been to 
write down a course grained free energy $F$ for the system and to write for 
Model A (non conserved dynamics)
\beq {\partial \rho \over \partial t} = - {\delta F \over \delta \rho} +
 \eta(x,t) \eeq
and for Model B (conserved dynamics)
 \beq {\partial \rho \over \partial t} =
 \nabla^2{\delta F \over \delta \rho} +\nabla \cdot\eta(x,t). \eeq
Here $\rho$ is not necessarily the density and is usually some scalar order
parameter for the problem, for example in a binary alloy it would correspond
to the difference between the densities of the two particle species and in
 a spin model it would be the local magnetization.
In Model B the noise is vector white noise of the type introduced above and
in Model A it is the scalar counterpart. The models  can lead to an 
equilibrium distribution for $\rho$ of the type that has as its free energy the
$F$ one started with, i.e. (formally) the equilibrium functional for $\rho$ is
given as
\beq P(\rho) = {1\over Z} \exp\lb -F[\rho]/T\rb \eeq
with $Z$ a normalizing factor. One may verify that the functional Fokker Planck
equation for equation (\ref{eq:prf6}) also has this as a steady state
solution. In this case it is interesting that the course grained free energy
has appeared directly and in addition without any appeal to there being a
large number of particles in the system, we recall that the term $\int dx 
\rho(x)\log\rho(x)$ normally occurs after a course graining and then a use of 
Stirling's formula. In this formulation we have avoided this route, presumably
it is the direct use of the Langevin process at the level of microscopic
dynamics that has smoothed out the statistics. However the functional 
representation of $P[\rho]$ does require additional reference to the underlying
physics when it comes to performing what is an {\em a priori} badly defined
functional integral. In the case of both Model A and Model B (and indeed for 
the statics) the usual tactic is to resort to an expansion about some fixed
background density up to fourth order in the fluctuations, this then yields
a dynamical version of Ginsburg--Landau theory. However we note that, even 
at the level of this approximation, equation (\ref{eq:prf6}) does not become
 equivalent to Models A or B (at this level of approximation one would still
be forced to keep terms of third order in the terms multiplying 
the white noise).

Once we have the full Langevin equation for the density evolution, we may use
the standard techniques of stochastic calculus (applied now to functions) in
order to obtain equations for the evolution of functions of the density. Here
we shall consider the time dependent correlation function at equal times. We
define
\beq C(x,y,t) = \langle \rho(x,t) \rho(y,t) \rangle.\eeq
It is straightforward to see that $C(x,y,t)$ obeys
$$ {\partial C(x,y,t)\over \partial t} =$$
\beq -T\nabla^2\delta(x-y)\langle\rho(x,t)\rangle +
2\langle\nabla \cdot\lb\rho(y,t)\rho(x,t) \int dz \rho(z,t) \nabla V(x - z)\rb
\rangle  + 
2T\nabla^2 C(x,y,t), \label{eq:prf8} \eeq
where we have use the symmetry between $x$ and $y$. Unfortunately the equation
(\ref{eq:prf8}) is not a closed equation for $C$. In general the above 
may be used to generate a whole hierarchy of equations for the $n$ point
functions of the field $\rho$, in fact this is the dynamical version of the
BBGKY hierarchy used in the static problem.

In conclusion we have presented a very simple derivation of the Langevin
equation of the density describing the dynamics of a gas of interacting
 Langevin processes.
In addition this formalism permits the construction of the BBGKY hierarchy
for the dynamical correlation functions in straightforward manner. Rather
than being a phenomenological model for the dynamics based around a known
course grained free energy, it is directly defined from an albeit simple
microscopic model. It has the merit of corresponding to a (at least 
in the sense of numerical simulations) realisable microscopic dynamics.
One rather appealing feature is that the naive course grained free energy
of the system appears naturally as the effective potential for the problem.

I would like to acknowledge useful discussions with L F Cugliandolo,
J Kurchan and G Parisi, and support from EU grant CHRX-CT93-0411.

\end{document}